# Clustering effects on wireless mobile ad-hoc networks performances

Gideon Naah[1] and Edwin Boadu Okoampa[2]

[1] School of Electronic Engineering, University of Electronic Science and Technology of China, Chengdu - China
*gnaah007@gmail.com*

[2] School of Micro-Electronics and Solid State Electronics, University of Electronic Science and Technology of China, Chengdu - China
*eddyboadu@gmail.com*

*Abstract*

*A new era is dawning for wireless mobile ad hoc networks where communication will be done using a group of mobile devices called cluster, hence clustered network. In a clustered network, protocols used by these mobile devices are different from those used in a wired network; which helps to save computation time and resources efficiently. This paper focuses on Cluster-Based Routing Protocol and Dynamic Source Routing. The results presented in this paper illustrates the implementation of Ad-hoc On-Demand Distance Vector routing protocol for enhancing mobile nodes performance and lifetime in a clustered network and to demonstrate how this routing protocol results in time efficient and resource saving in wireless mobile ad hoc networks.*

*Keywords*

AODV, CBRP, DSR, MANETs, Routing.

## 1. Introduction

A wireless mobile ad-hoc network is a group of mobile devices that forms a network which does not require the usage of wires or cables for communication. Mobile nodes in this network are able to detect the presence of nodes that are in close proximity. Due to the limited transmission range of wireless network interfaces, multiple networks "hops" may be needed for one node to exchange data with another across the network. Wireless ad-hoc networks have some properties such as the dynamic network topology, limited bandwidth and energy constraint in the network as described by Kumar et al [1]. Mobile ad hoc network (MANET) is useful for different purposes e.g. military operation to provide communication between squads, collaborative and distributed computing, wireless mesh control, wireless sensor networks, hybrid network, medical control etc. Kumar et al [1] said "routing protocol plays very important part in implementation of mobile ad hoc networks. The following are the main protocols used in routing:

Proactive or table driven routing protocols and Reactive or on-demand routing protocols. DSR (Dynamic Source Routing) requires no periodic packets of any kind at any level within the network as stated by Johnson et al [2].

DSR does not use any periodic routing advertisement, link status sensing, or neighbour detection packets, and does not rely on these functions from any underlying protocols in the network. This entirely on-demand behaviour and lack of periodic activity allows the number of overhead packets caused by DSR to scale all the way down to zero, when all nodes are approximately stationary with respect to each other and all routes needed for current communication have already been discovered also as stated by Johnson et al [2]. As nodes begin to move more or as communication patterns change, the routing packet overhead of DSR automatically scales to only that needed to track the routes currently in use. In response to a single Route Discovery (as well as through routing information from other packets overheard), a node may learn and cache multiple routes to any destination. This allows the reaction to routing changes to be much more rapid, since a node with multiple routes to a destination can try another cached route if the one it has been using should fail. This caching of multiple routes also avoids the overhead of needing to perform a new Route Discovery each time a route in use breaks. The operation of Route Discovery and Route Maintenance in DSR are designed to allow unidirectional links and asymmetric routes to be easily supported. In wireless networks, it is possible that a link between two nodes may not work equally well in both directions, due to differing antenna or propagation patterns or sources of interference as described by Johnson et al [2]. DSR allows such unidirectional links to be used when necessary, improving overall performance and network connectivity in the system [1].

DSR also supports internetworking between different types of wireless networks allowing a source route to be composed of hops over a combination of any types of networks available by Johnson et al [2]. For example, some nodes in the ad hoc network may have only short-range radios, while other nodes have both short-range and long-range radios; the combination of these nodes together can be considered by DSR as a single ad hoc network [2].

DSR is an on-demand routing protocol and cannot perform well in a large MANET and the reason is that it has scalability issues when the size of the network increases, mostly when there is node mobility simultaneously.

Proactive routing requires control overhead for building and updating this table, having information about the state of the network. For on-demand routing protocols, routes are found when required; however this causes it to suffer significant route setup delay which becomes intolerable in the presence of both a large number of nodes movement. The fundamental idea of on-demand routing protocols is that, an initial node sends a route request and makes a decision based on the reply received, which may be sent by an intermediate mobile node. However, on-demand routing algorithms have the disadvantage of increasing per-packet overhead. This per-packet overhead reduces the available bandwidth for information transmission in the network. Due to the routine of dissemination of path requests (flooding), it is difficult to reduce the dissemination of packets unnecessarily [1]. Management of large number of nodes is one of the essential issues ad-hoc network faces.

The nodes of a wireless ad-hoc network are divided into numerous fragmented or intersecting clusters. Each cluster elects one node as the so-called cluster head and these distinct nodes are accountable for the routing process. Neighbours of cluster heads cannot be cluster heads as well.

But cluster heads are able to communicate with each other by using *gateway nodes*. A gateway is a node that has two or more cluster heads as its neighbours or— when the clusters are disjoint—at least one cluster head and another gateway node.

Cluster heads helps in the reduction of traffic and this is because requests will only be sent through them. Information transmission is therefore done only by these cluster heads from one cluster to the other. As in any other categorized protocol used in routing, there are overheads associated with cluster formation and maintenance as described by Sk. Munwar and Dr. V.V.Rama Prasad [3].

**Figure 1: A cluster based Ad Hoc network: [4]**

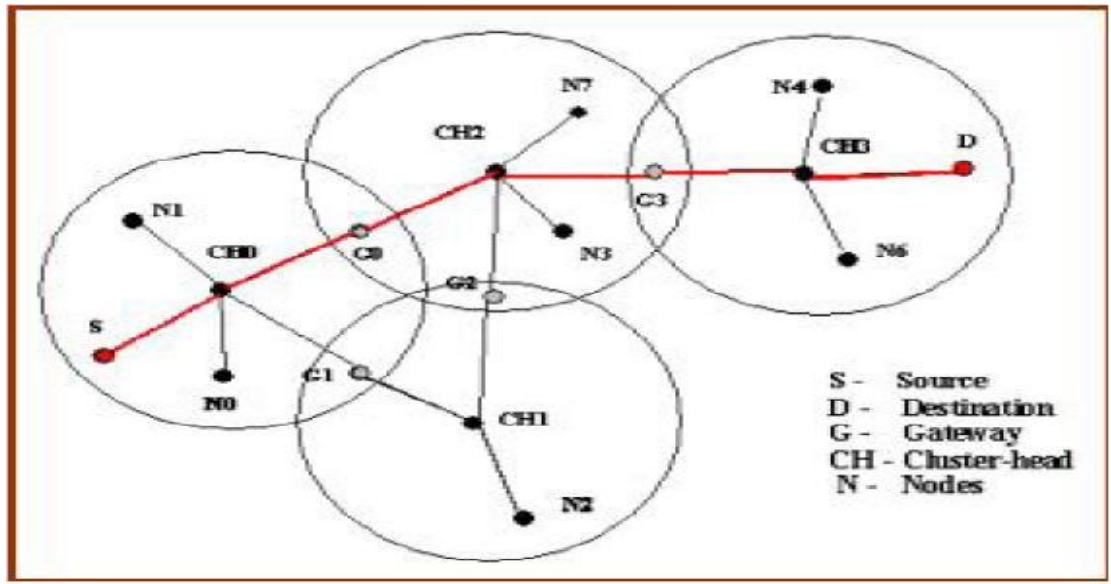

Pattern used by CBRP:
Source → Cluster head → Gateway → Cluster head → Gateway → ⋯ → Destination

This paper presents the implementation of *Ad-hoc On-Demand Distance Vector (*AODV) routing protocol to enhance the performance and lifetime of nodes in a clustered network and to demonstrate how this routing protocol results in time efficient and resource saving in mobile ad hoc networks. AODV is implemented using Network Simulator-2 (NS-2) which is the simulation environment. NS-2[5] is an object-oriented discrete event driven network simulator developed at University of California (UC) Berkeley written in C++ and Object Tool Command Language (OTcl).

## 2.0 Related works

### 2.1 Ad-Hoc Network

Toh [6] described an ad-hoc network as "a collection of two or more devices equipped with wireless communications and networking capability." Communication in ad hoc network can be done directly between nearby nodes in close proximity or use an intermediate node to reach nodes that other nodes cannot reach.
There are two main merits ad hoc networks can boast of and they are as follows:
(a) Has no wired/cable connected systems put in place
(b) Has the ability of creating its own network easily
That is in a conventional network, devices like switches, cabling/wiring, base-stations, and routers are not required. A deployment strategy was proposed to increase the probability of a 'reliable path'. No other protocol has been developed to efficiently enhance the performance and lifetime of nodes in ad-hoc network.
Ad hoc network's issues of design was introduced by [7], summarizing an ad-hoc networking design as follows: "operate each mobile node as a specialized router, which periodically advertises its view of the interconnection topology with other mobile nodes within the network." The exceptional part here is dealing with the effects of nodes mobility. That is, ad-hoc networking protocols must be able to sense and then react to changes in the network's environment. Freebersyser and Leiner [8] described the implementations of the MANET for Department of Defence (DoD) applications.

## 2.2 Cluster-based MANET

The clustered MANET is an extension of the ubiquitous MANET schema. As the size of the network (number of mobile nodes) increases, resources such as bandwidth becomes limited. Clustering however, is able to break the network into groups which makes dissemination of packets easily and at the long run; it enhances and widens the usage of resources that were previously limited. This is done by dividing the ad hoc network into numerous smaller sub networks. This sub networks are then brought together by what is termed back-bone network. Numerous nodes are then selected as the back-bone nodes. Together they form the back-bone network. A popular network design includes two back-bone networks.
The use of cluster based MANET for wireless sensor networks were also addressed. In the literatures, the researchers developed a Medium Access Control (MAC) scheme that is event driven.

However, clustering consumes bandwidth, drain mobile nodes energy quickly, causes congestion, collision and data delay in larger networks because of beacon messages i.e. periodic hello messages been broadcast at every interval.

Some simulation results for DSR, Destination Sequenced Distance Vector (DSDV) and Temporarily Ordered Routing Algorithm (TORA) have been presented in earlier papers, those simulations used considerably different input parameters which differ from this paper and did not simulate the wireless network as accurately. Currently, there is no awareness of any previously published performance enhancement results for mobile nodes using AODV in wireless mobile ad hoc networks environments.

## 2.3 TORA

Some researchers named Park and Corson, also worked on another protocol named Temporarily Ordered Routing Algorithm (TORA) as named above. In order to avoid congestion, their simulation used a packet transmission rate of only 4, 1.5, or 0.6 packets per *minute* per node. Their work showed that the nodes were divided by links and these nodes had two states namely active state and inactive state. Active state means that the node is operational and inactive state also means that the node is not operational. These links were error free and no dynamics below the network layer were modeled.

## 2.4 Johnson and Maltz

Johnson and Maltz previously simulated DSR [9] using the same mobility model as in other papers. Their work had issues with the MAC layer and the way its radio was propagated. Furthermore, broadcast and unicast packets were delivered with the same probability and this is not a realistic assumption.

## 2.5 Freisleben and Jansen

Freisleben and Jansen [10] did a simulation and a comparison on DSR and DSDV. Their simulations used configurations of 10 mobile nodes, with movement speeds relative to transmission range, making some of their results unrealistic. All events in their simulator took place in regular time steps, resulting in perfectly synchronized behaviour of a number of separate mobile nodes in some cases.

## 3.0 Methodology used

Recently, a diversity of different and novel routing protocols have been developed to study the movement of nodes in wireless mobile ad hoc networks, but employing a mechanism for enhancing mobile nodes performance and lifetime in a larger and dense clustered networks and to demonstrate how this routing protocol results in time efficient and resource saving in mobile ad-hoc networks is not yet available. This paper adaptively optimizes the frequent needs of those messages using AODV protocol. The proposed AODV algorithm used to model the uncertainties for updating local connectivity successfully in time. AODV routing algorithm was developed from two other routing protocols namely DSDV (Destination Sequenced Distance Vector) and DSR routing algorithms. The AODV protocol took two properties from the DSDV routing algorithm and one property from the DSR routing algorithm. These properties are as follows**:**

1) The Sequence Numbering Process and Periodic Hello Messaging from the DSDV routing protocol.

2) The Route Discovery Process from the DSR routing protocol.

The following describes how the AODV protocol operates within the wireless network:

**FIGURE 2 [11]**

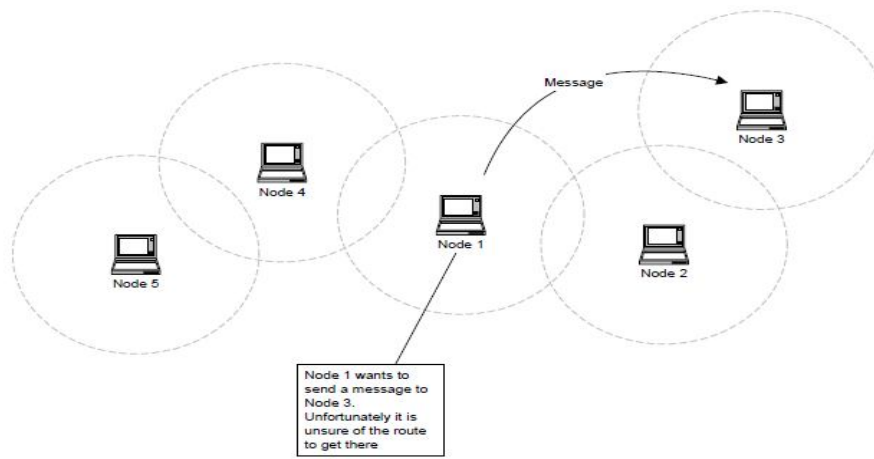

AODV is a technique of routing messages between mobile devices. It allows these mobile devices or nodes to transmit messages through their neighbours to nodes they cannot communicate with directly. These nodes do not keep or maintain routing tables but rather stores information about the network in a form of pointers for easy referencing. AODV is able to do this by discovering the routes along which messages can be transmitted. This protocol sees to it that paths that are required for transmitting and receiving data does not repeat itself over and over again termed loop and tries to find the most shortest path if the need be. When there are alterations in paths required for transmitting and receiving data, and if there are errors in these paths; this protocol is also able to handle all these. This can be seen as shown in the figure above.

**FIGURE 3 [11]**

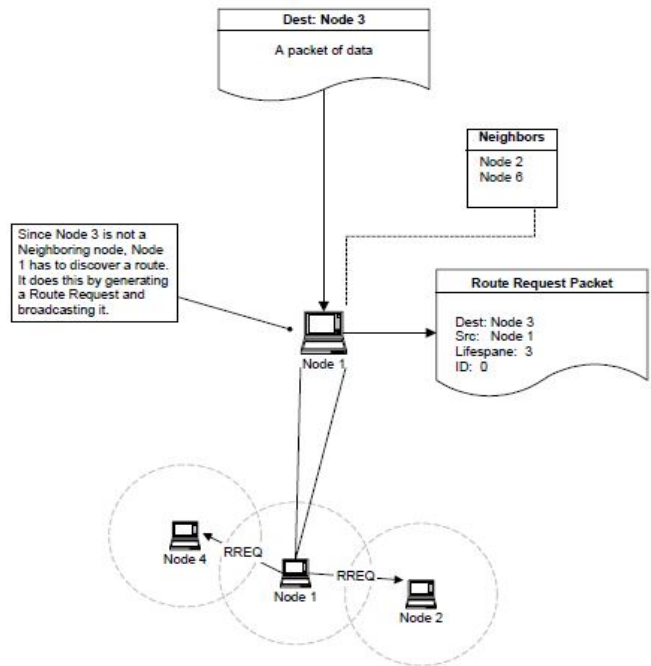

When nodes have a bidirectional link with one other, which means that they can communicate with each other directly; they are referred to as neighbours. For a node to know the location of other nodes (neighbours), it listens to messages broadcasted at certain time intervals and it uses this process to track its neighbours. For a node to send a message to another node that is not in close proximity, it transmits what is called a Route Request (RREQ) message. This message has several kinds of information namely: the source, the destination, the lifespan of the message and a Sequence Number which serves as a unique ID.

In the example as shown in Figure 3 above, Node 1 wishes to send a message to Node 3. Node 1's Neighbours are Nodes 2 and 4. Since mobile node 1 cannot directly communicate with mobile node 3, mobile node 1 sends out a RREQ. The RREQ is heard by mobile nodes 2 and 4.

**FIGURE 4 [11]**

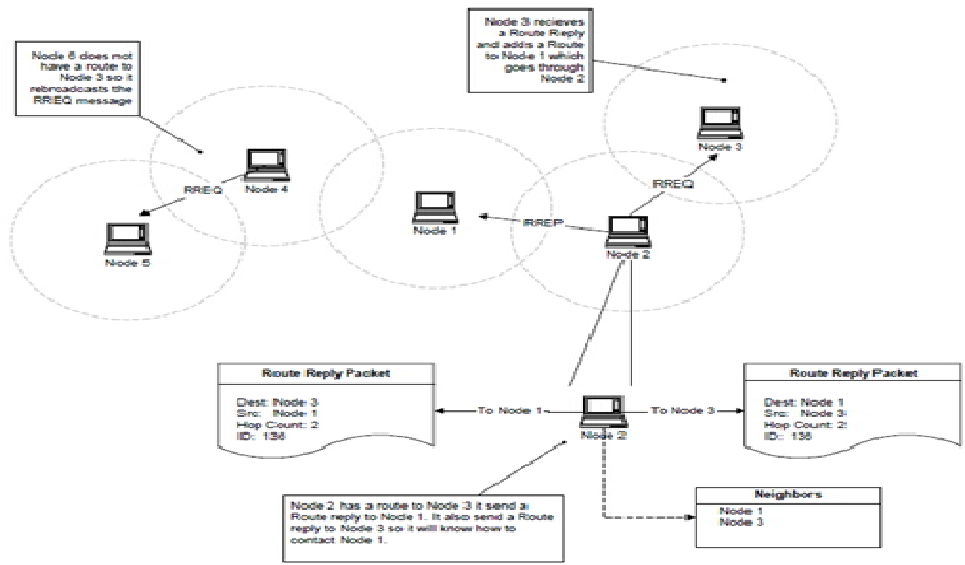

For neighbours of mobile nodes that are not in close proximity such as neighbours of mobile node 1 receives a message, they check to see whether the message is meant for them or they know the best path to transmit this message to the rightful owner. If the message is meant for them, they will just send back a reply message to node 1; other than that will retransmit the request message to their set of neighbours. This message keeps getting retransmitted until its lifespan is over.

If node 1 does not receive a reply in a certain time interval, it will retransmit the request except this time the request message will have a longer lifespan than previous and a new ID number. These mobile nodes uses what is called Sequence Number in the request message to insure that they do not retransmit a request message. This can be seen in the figure above as shown.

**SEQUENCE NUMBERS**

**FIGURE 5 [11]**

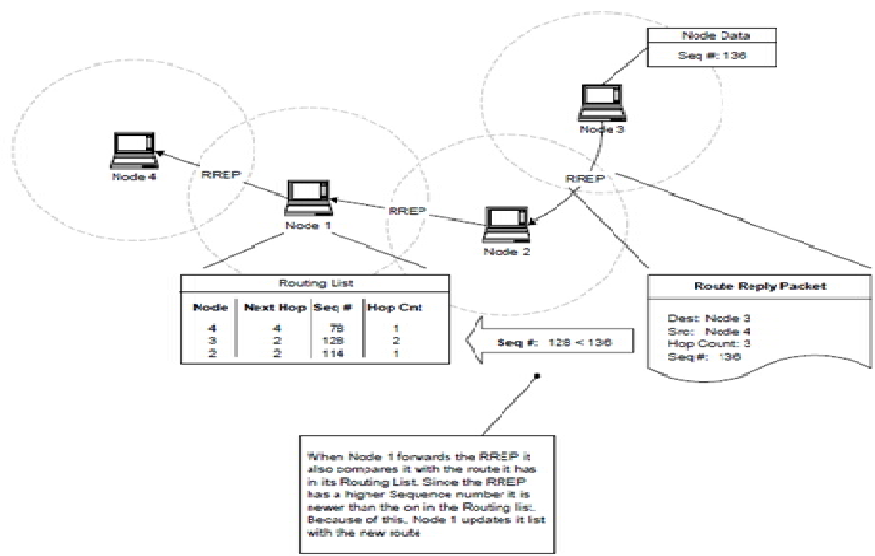

Sequence numbers serve as time stamps. They allow mobile nodes to compare how current their information on other nodes are. A mobile node always increases its sequence number any time it transmits information/data. Every time a mobile node communicates with its neighbours, it keeps records of sequence numbers belonging to all the mobile nodes it had interacted with. When the sequence number is high, it tells that that is the current and newer path which can or must be used for transmitting and receiving information. In other words, a higher Sequence number signifies a fresher route. More accurate and current information can be obtained as a result of using this process.
This can be seen as shown in the figure above.

**ERROR MESSAGES**

**FIGURE 6 [11]**

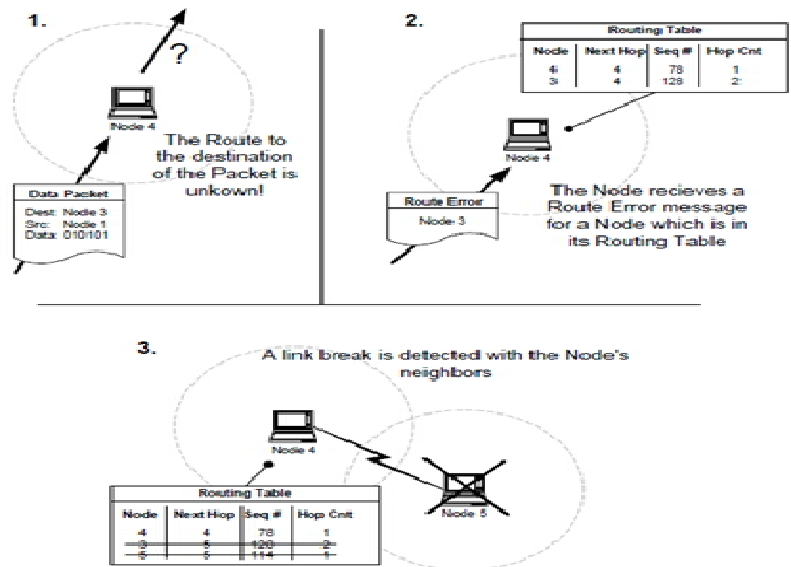

For AODV to adjust and make updates to changes in paths that are required for communication, it uses what is called the route error message (RERR) in doing this. It uses this RERR message in omitting/deleting mobile nodes paths that are irrelevant and no more in use or required, and it does this by using its referencing point.
There are situations whereby these mobile nodes cannot communicate with one or more mobile nodes and when this happens, it looks at its referencing point for Route that uses the Neighbour for a next hop and marks them as invalid. Then it sends out a RERR with the Neighbour and the invalid routes.
This can also be seen in the figure above.

## 4.0 EXPERIMENTAL RESULTS

**FIGURE 7: CONSOLE MODE IN RED HAT ENTERPRISE LINUX SERVER 5 OPERATING SYSTEM**

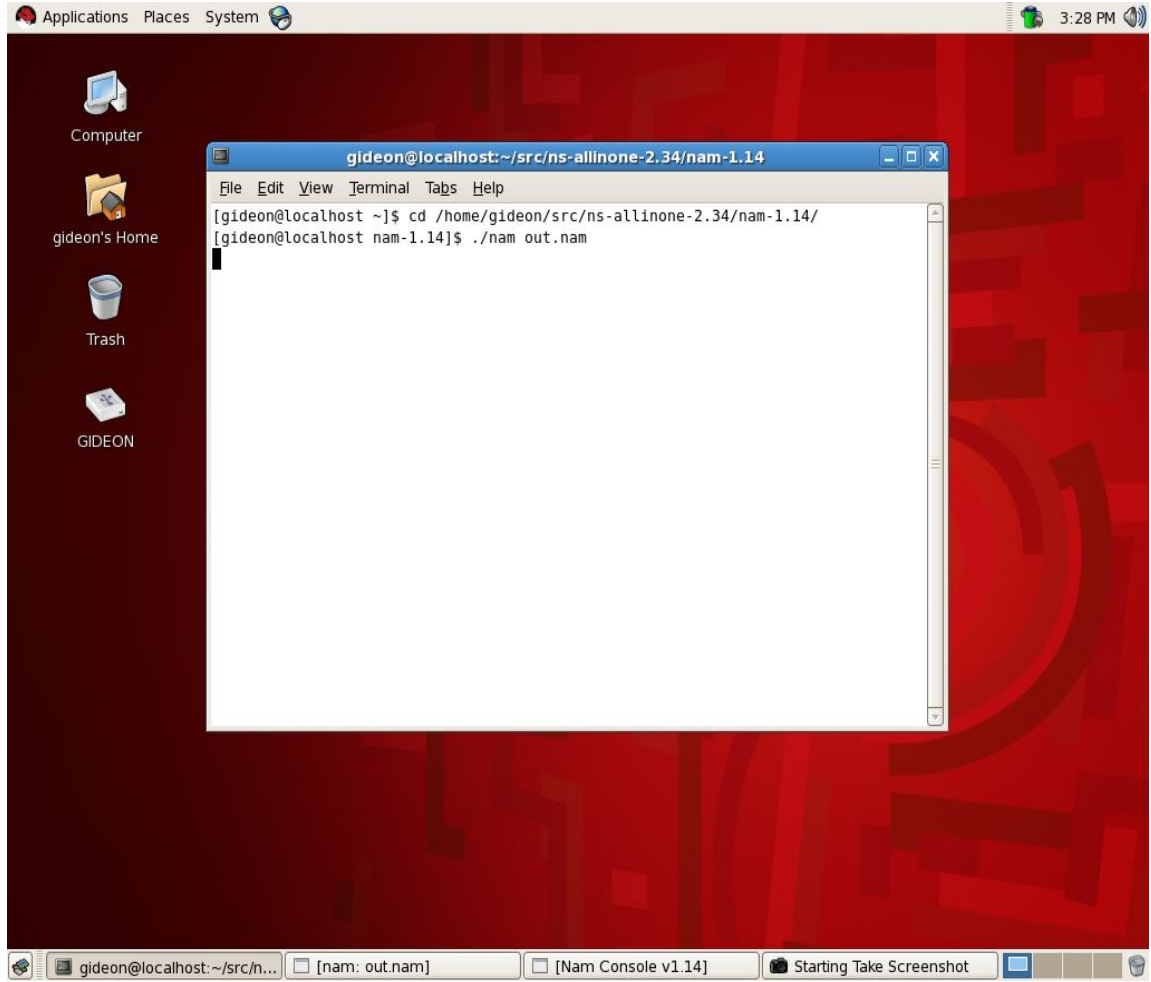

Figure 7 above shows the console mode of Red Hat Enterprise Linux Server 5 Operating System Edition where the NS-2 applications and Tcl codes can be run for the required animator called NAM in NS-2 to be displayed.

**FIGURE 8: SHOWING NAM (NETWORK ANIMATOR) MEANT FOR THE SIMULATION PROCESS**

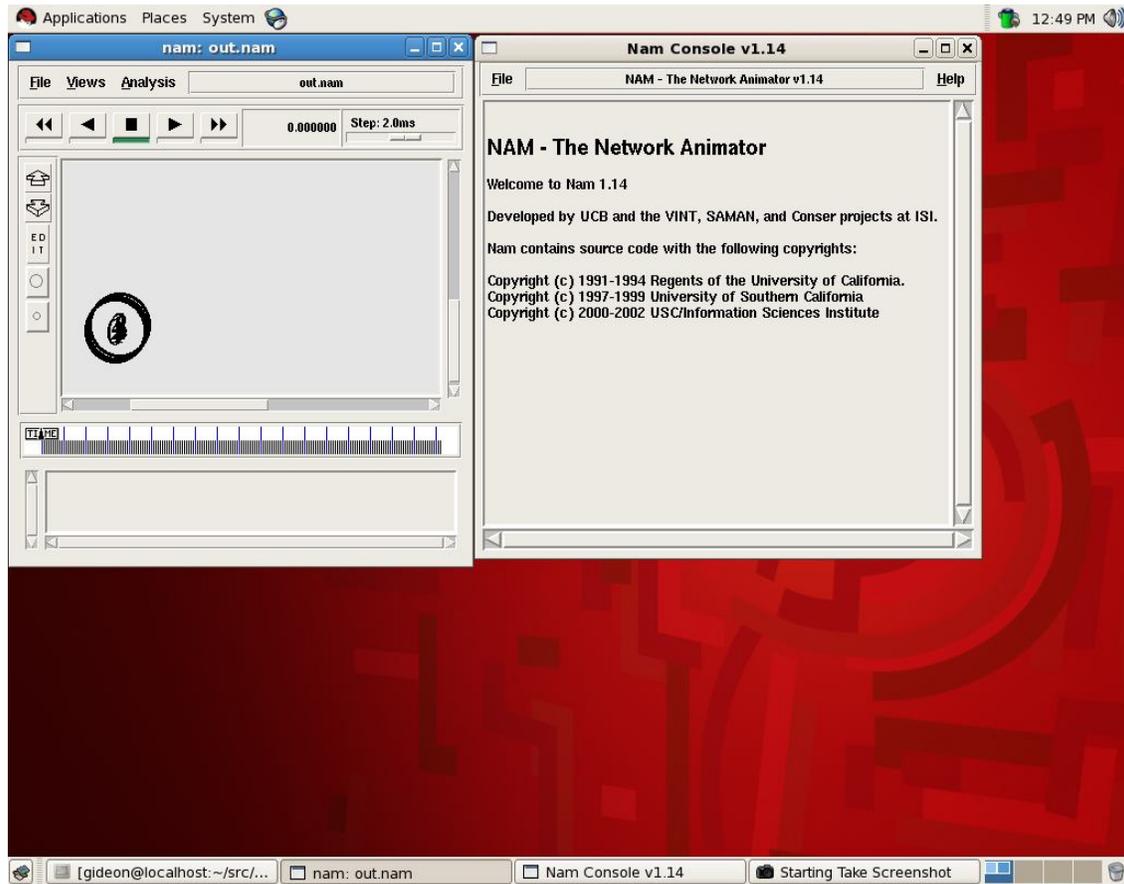

In Figure 8 above, this is where the simulation of the AODV routing protocol will take place between the nodes. The nodes in the nam are what to be used for the simulation. The nodes are at their initial positions as shown above and they are 5 in all. Five nodes were used for clearer view of the simulation.

**FIGURE 9: NAM WITH NODES MOBILITY**

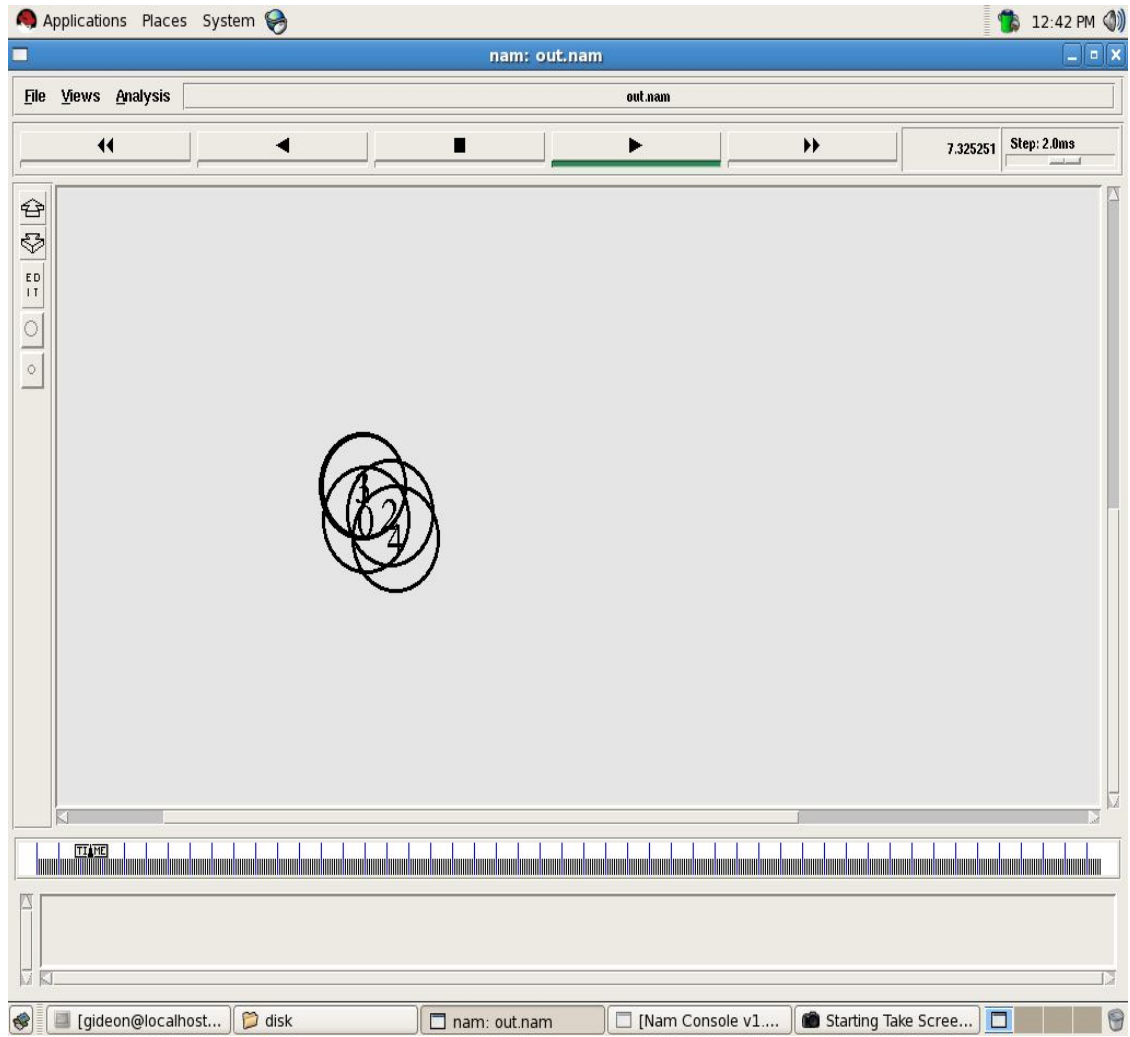

In Figure 9 above, this is where mobile nodes within the wireless ad-hoc network began the clustering process i.e. organizing themselves to form a cluster before communication takes place between them. The 5 nodes in the nam have now completely formed a cluster at time 7.325251 seconds and each node therefore knows the location of its neighbours'. Therefore sending data packets to any other node would not cause any difficulty. Communication of data packets among these mobile nodes took place mainly between nodes 1, 0 and 4.

**FIGURE 10: NAM DISPLAYING THE SIGNALS EACH MOBILE NODE IS TRANSMITTING WITHIN THE WIRELESS AD-HOC NETWORK AND ALSO SHOWING THAT COMMUNICATION IS STILL POSSIBLE THOUGH THERE IS CONGESTION WITHIN THE WIRELESS AD-HOC NETWORK.**

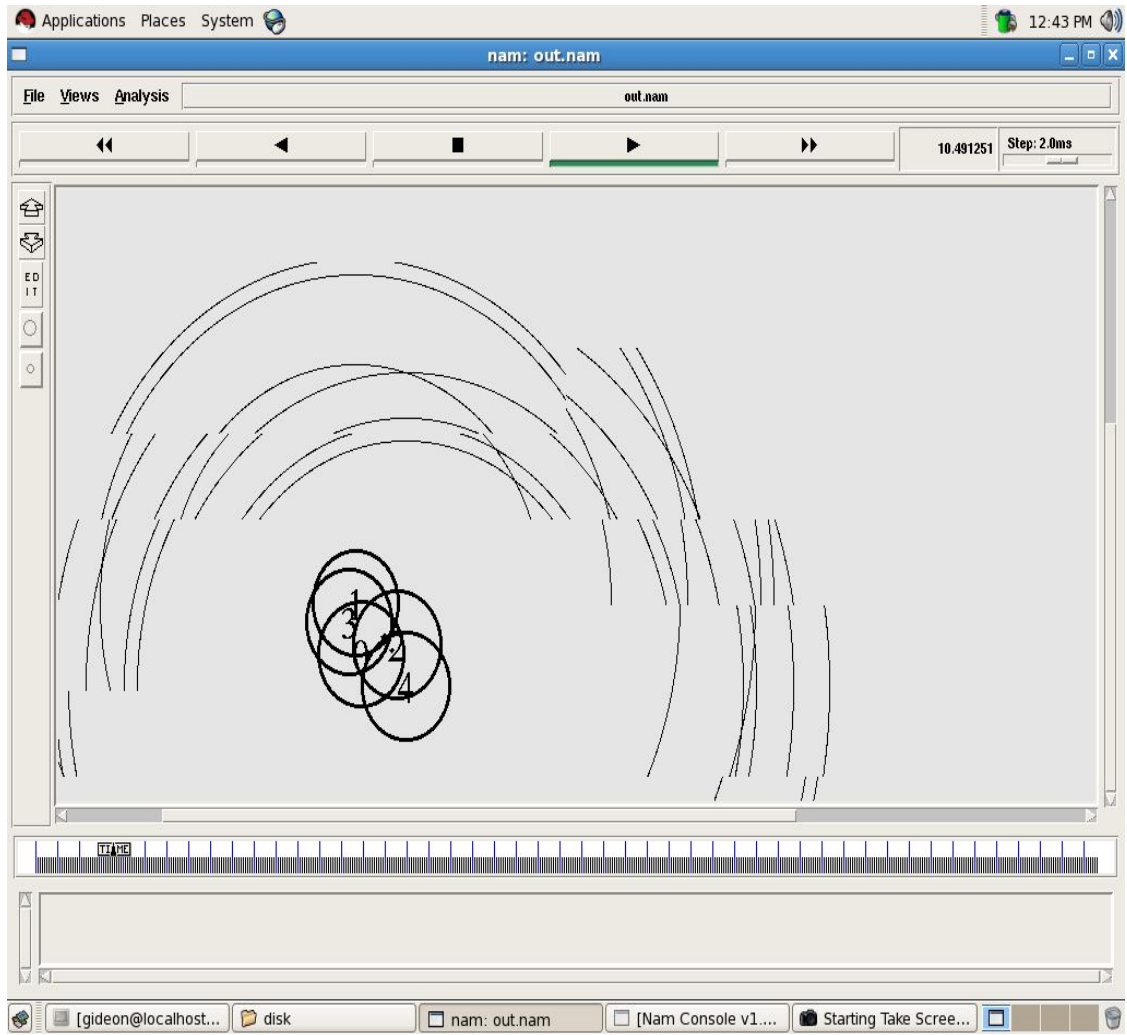

In Figure 10 above, each node within the wireless ad-hoc network is transmitting a signal in order to sense the presence of its members before communication takes place. After connection is established between the wireless mobile nodes, any data packets can now be transmitted between these mobile nodes. In the diagram, communication is still taking place between the nodes such as nodes 1 and 4 though these mobile nodes are congested. The routing process i.e. choosing the best path to transmit and receive a message by mobile nodes began at time 10.491251 seconds. All the nodes are overlapping at this stage for easy transfer of data packets. When it gets to a point where a node is out of its original position i.e. out of its coverage area, intermediate nodes are able to relay data packets on behalf of other nodes and this is one of the main purposes of ad-hoc network.

**FIGURE 11: NAM SHOWING THAT MOBILE NODES LIFETIME IS ENHANCED WHEN NODE 1 WENT OUT OF ITS RANGE TOTALLY AND THERE IS STILL COMMUNICATION BETWEEN NODES 1 AND 4.**

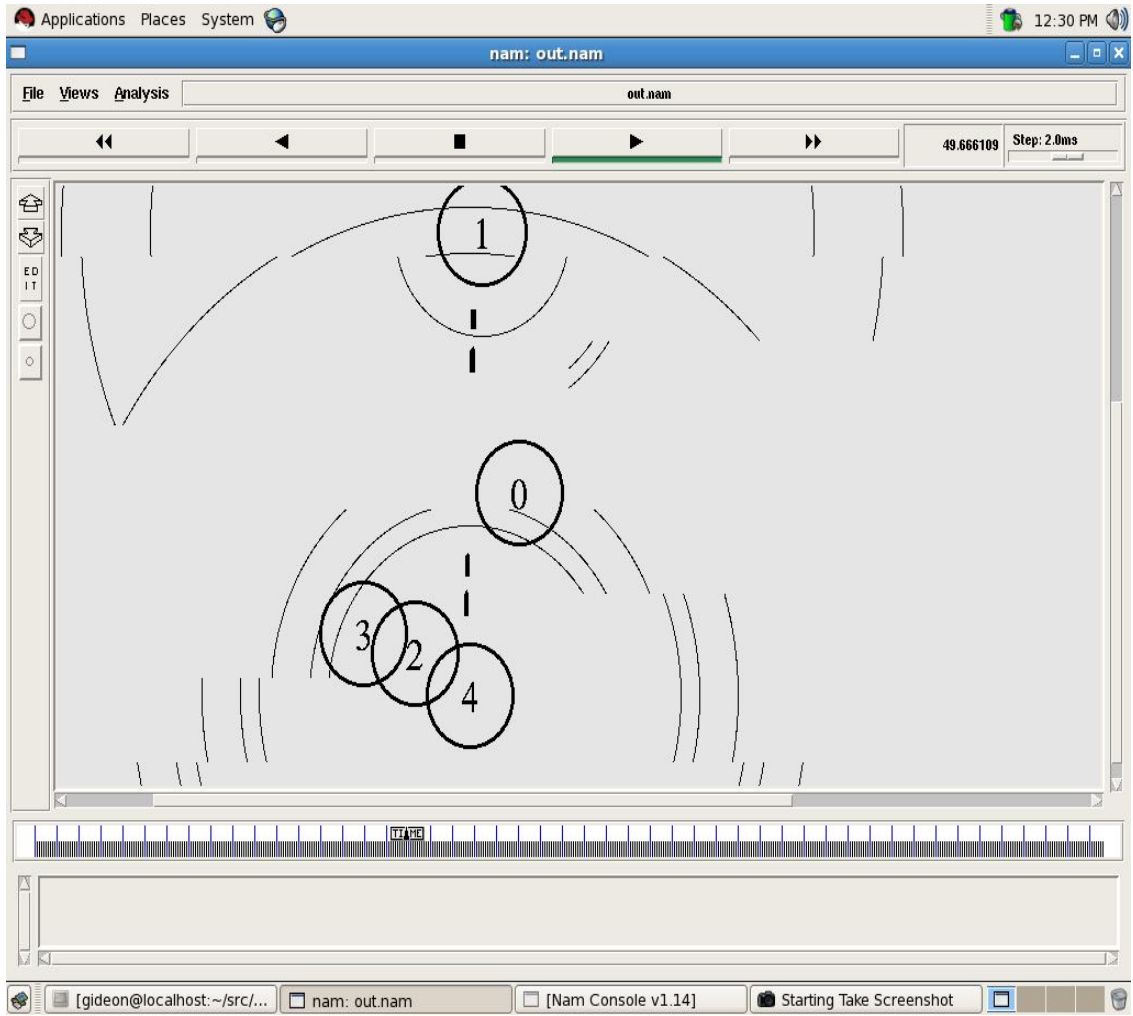

Figure 11 above shows that the mobile nodes 3, 2, and 4 are now overlapping each other while nodes 1 and 0 are separate nodes and also out of their initial position. This occurred at time 49.666109 seconds. Therefore, lifetime in a wireless ad-hoc network has been enhanced though node 1 is out of its coverage area and there is still communication between nodes 1 and 4. When it happens this way, the node sending the data packets i.e. node 4 uses so much energy in transmitting this message to node 1 just to ensure that node 1 receives the message. Unless node 1 is out of the network before it cannot receive the message. While this was going on, node 0 was also moving due to frequent mobility that takes place in the wireless environment. The distances of nodes 1 and 0 are relative to node 4.

**FIGURE 12: NAM DISPLAYING PACKETS DROPPED WHEN NODE 1 WENT OUT OF ITS COVERAGE AREA**

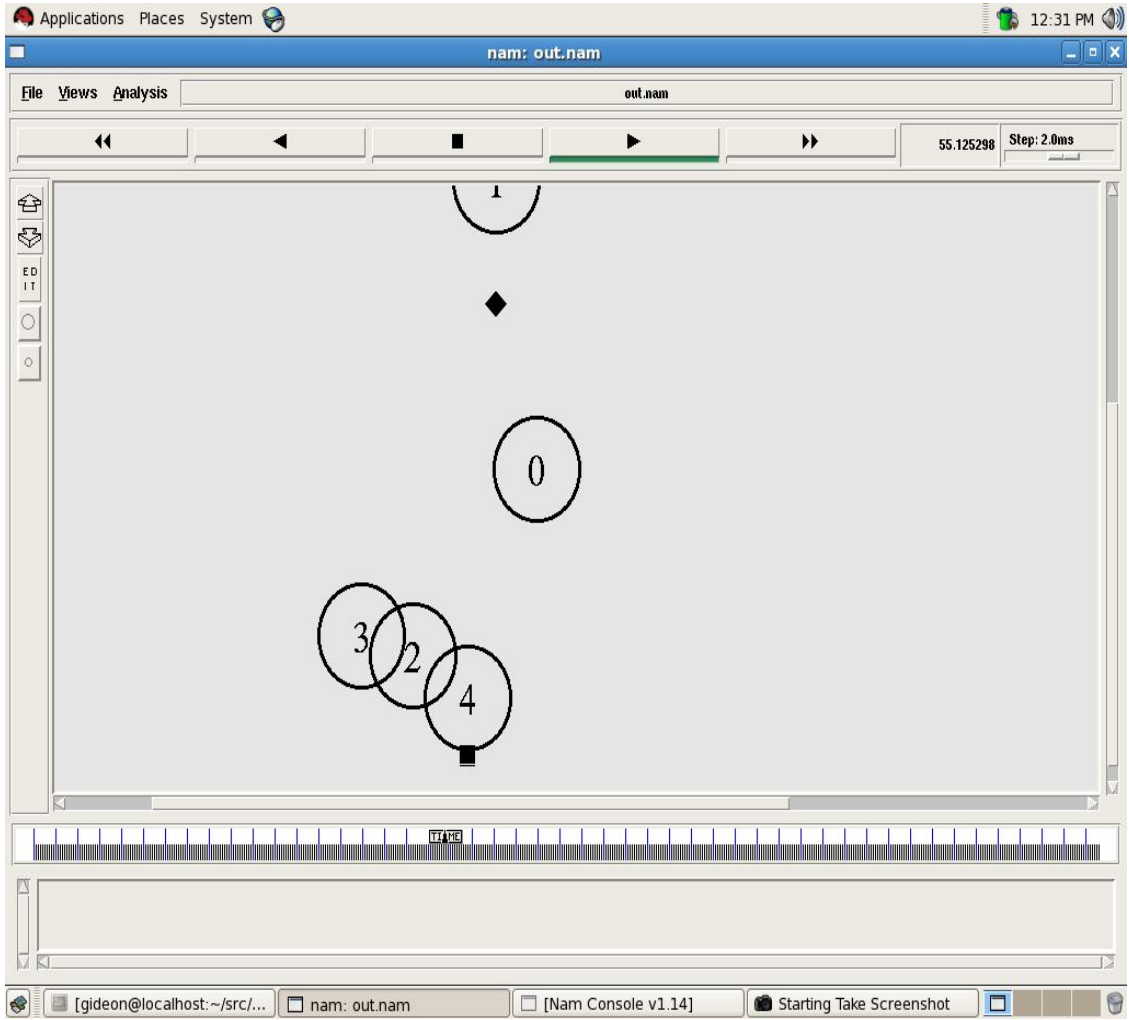

In Figure 12 above, nodes 1 and 4 are no more communicating and as a result both nodes dropped their data packets. The squared-like structures falling from nodes 1 and 4 are the data packets been dropped or lost. This happened due to mobility of nodes within the wireless environment. This is because node 1 changed its original position and went further away from its radius. Node 4 then uses its point of reference to quickly look for nearby nodes who can relay data packets on its behalf. Node 4 then finds the nearest neighbour node which is node 0 which finally relayed its data packets to node 1 which were communicating initially.

**FIGURE 13: NAM SHOWING THAT COMMUNICATION IS STILL TAKING PLACE BETWEEN NODES 1 AND 4.**

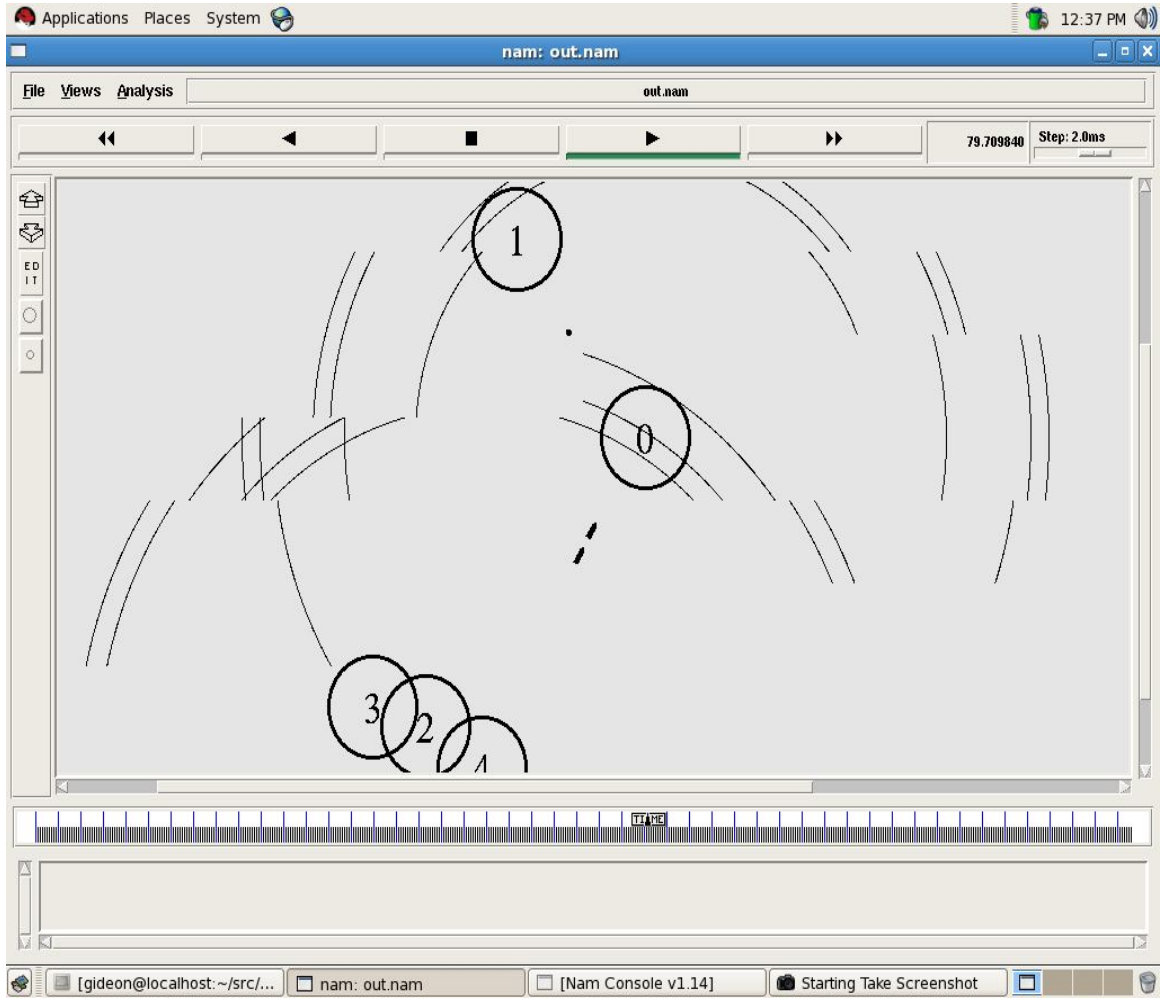

Figure 13 above shows that though node 1 went further away from its coverage area due to mobility of wireless mobile nodes in ad-hoc network, node 4 quickly looks at its referencing point and uses the nearest node i.e. node 0 to still relay packets to node 1. In this way, node 0 transmits packets on-behalf of node 4. This process also took place at time 79.791880 seconds. However, it uses sequence numbers to indicate how fresh or current the information is within the network. When a mobile node receives a message, it checks to see if the message is meant for it. If the message is intended for it, it then sends a reply back to the source and if not; it sends the message to other nodes within the network. In this case, the node checks the sequence number within the message received and if it is higher than what it has, it automatically increases or updates its own to the current one before either sending or accepting the message. The nodes also do not maintain routing tables but rather stores information about the network in a form of pointers for easy referencing.

**FIGURE 14: XGRAPH ANIMATOR DISPLAYING END OF SIMULATION RESULTS.
GRAPH: DISTANCE OF MOBILE NODES IN METRES(M) ON (Y-AXIS) AGAINST TIME
IN SECONDS(SEC) ON X-AXIS.**

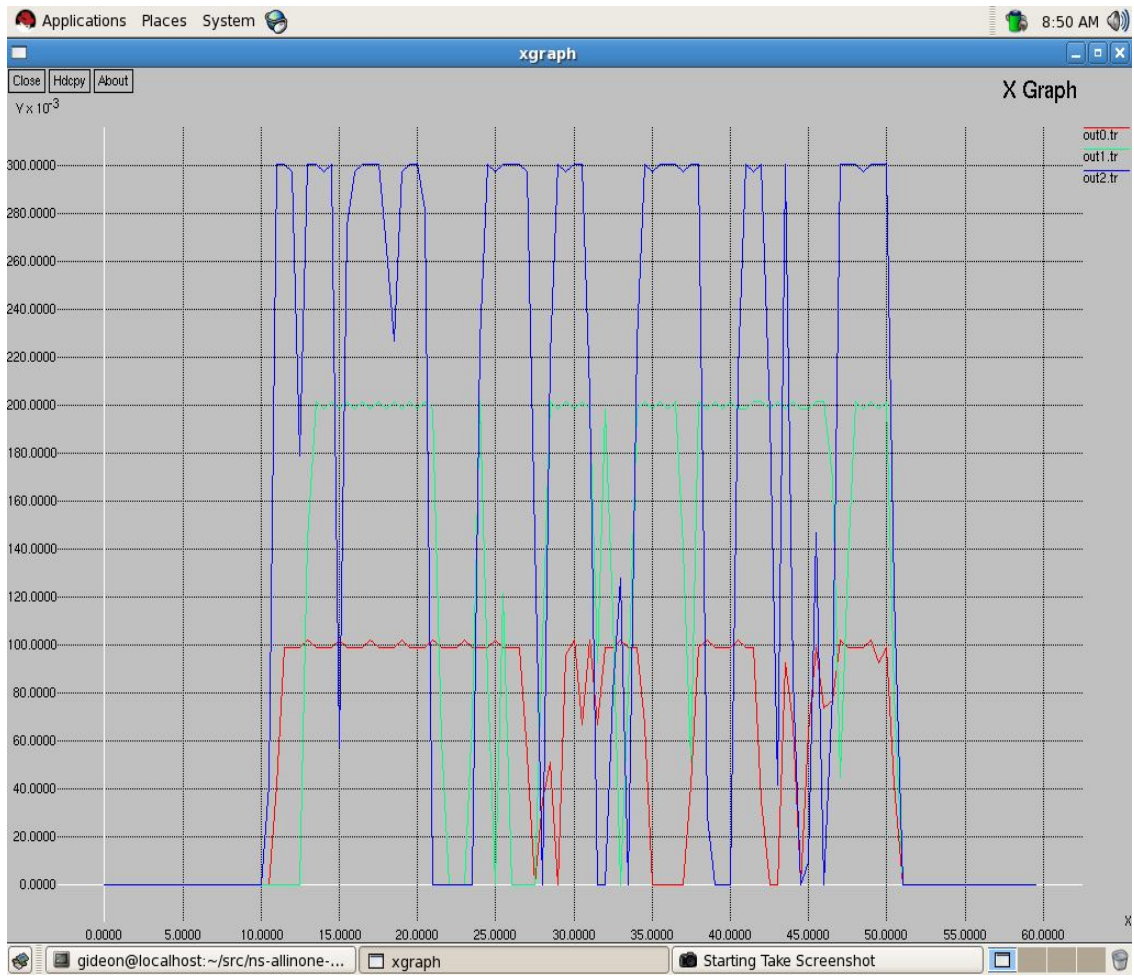

Xgraph is a graph plotting software in NS-2 that plots a graph and displays the results after simulating the Tcl script. After simulating the tcl script, the output of the tcl script is then sent as input to the xgraph which then plots the graph for the simulation based on this input. From the graph, the Y-axis stands for the distances of mobile nodes 0, 1, 4 and the X-axis stands for the time used by these nodes for the simulation in seconds. The distances of these mobile nodes i.e. nodes 0 and 1 are relative to node 4. These were the main nodes used for communication within the simulation. The clustering process began at time 0.0000seconds and the routing process also at time 10.0000seconds. The highest peak which is blue in colour indicates the node mobility of node 1. As node 1 moved further away from its coverage area, it is still possible to receive data packets from Node 4. The next peak in rank which is also green in colour indicates the node mobility of node 0. It also moved further away from its coverage area as node 1 moves. Node 0 is therefore able to transmit data packets on behalf of node 4. This is the purpose of ad-hoc networks i.e. intermediate nodes should be able to forward or relay data packets on behalf of other nodes that are totally out of its transmission range.
Finally, the last peak which is red in colour indicates node 4 which also moved slightly away from its initial position in order to still make communication possible between itself and node 1. The simulation ended at time 60.0000seconds.

The rise in peak indicates the point at which mobile nodes are able to receive data packets at a higher rate and the fall in peak indicates the point at which mobile nodes are still able to receive data packets but at a lower rate due to mobility of nodes within the wireless environment. Certain environmental factors such as rain and foliage can also limit the transmission of data packets.

**Table 1: routing time for mobile nodes from nam**

| MOBILE NODES DATA TRANSFER | AODV ROUTING TIME (SECONDS) |
|---|---|
| ALL NODES | 7.565666 - (CLUSTERING PROCESS) 10.777777- (ROUTING PROCESS) |
| BETWEEN NODES 1 AND 4 | 49.999999 |
| BETWEEN NODES 1, 0 AND 4 | 79.989898 |

## 5.0 CONCLUSIONS

A lot of research had been conducted within the wireless mobile ad hoc network environments these recent years using numerous protocols. Mobile Ad hoc network has a brighter future and more prospects to deliver. A lot of computing devices are being developed day in day out with very high processing speeds at a faster rate. However, not much have been said or done for clustering systems in mobile ad hoc networks using AODV protocol. This paper presents on such kind. The study conducted in this paper reveals that the AODV protocol does not maintain any routing tables in the nodes but rather stores information about the network in a form of pointers for easy referencing which results in less overhead and more bandwidth availability. The AODV routing protocol also consumes less bandwidth, decreases overhead and able to support very large number of mobile nodes in a congested and clustered network. As a result, it enhanced mobile nodes life time and performance in clustering systems within MANETs.

The challenges faced by the AODV routing protocol used within the clustering system in a wireless MANET environment are as follows**:**

Mobile nodes that are transitional can lead to inconsistent routes if the source sequence number is very old and obsolete; and these transitional nodes may have a higher but not the latest destination sequence number, thereby having out-of-date entries. Secondly, multiple Route Reply packets that is in response to a single Route Request packet can lead to a heavy control overhead.

When data transfer is taking place between mobile nodes and one of the nodes changes its position, the data packets been transferred are loss. The node sending the data packets has to use its referencing point in a form of pointer to look for the immediate neighbour node quickly that can relay data packets on its behalf to that particular node it was communicating with earlier on. The loss of data packets are normally triggered either by mobility of mobile nodes or environmental factors such as rain or foliage. The loss of data packets between mobile nodes 1 and 4 as shown in Figure 12 above occurred at time 55.125298seconds. Further study will be focused on solving AODV's routing protocol limitations for clustered networks in MANETs.